\documentclass{iopart}
\usepackage{iopams}

\usepackage{txfonts}
\usepackage{amssymb,amstext,amsfonts}
\usepackage[final]{graphicx}
\usepackage[sort&compress,comma,square,numbers]{natbib}
\usepackage[draft]{hyperref}

\renewcommand\vec[1]{\boldsymbol{\mathrm{#1}}}
\newcommand\diff{\mathrm{d}}

\newlength\figurewidth
\AtBeginDocument{\setlength\figurewidth{.65\linewidth}}

\begin{document}
\title{Anomalous transport of a tracer on  percolating clusters}
\author{Markus Spanner\textsuperscript{1},
Felix H\"ofling\textsuperscript{2},
Gerd Schr{\"o}der-Turk\textsuperscript{1},
Klaus Mecke\textsuperscript{1}, and
Thomas Franosch\textsuperscript{1}}
\address{\textsuperscript{1}Institut f\"ur Theoretische Physik,  Universit\"at Erlangen-N\"urnberg, Staudtstra{\ss}e~7, 91058 Erlangen, Germany \\
\textsuperscript{2}Max-Planck-Institut f\"ur Metallforschung,
Heisenbergstra{\ss}e 3, 70569 Stuttgart and Institut f\"ur Theore\-tische und Angewandte Physik, Universit\"at
Stuttgart, Pfaffenwaldring 57, 70569 Stuttgart, Germany}

\date{\today}

\begin{abstract}
We investigate the dynamics of a single tracer exploring a course of fixed obstacles in the vicinity of the percolation transition for particles confined to the infinite cluster. The mean-square displacement displays anomalous transport, which extends to infinite times precisely at the critical obstacle density. The slowing down of the diffusion coefficient exhibits  power-law behavior for densities close to the critical point and we show that the mean-square displacement fulfills a scaling hypothesis. Furthermore, we calculate the dynamic conductivity as response to an alternating electric field.
Last, we discuss the non-gaussian parameter as an indicator for heterogeneous dynamics.
\end{abstract}




\maketitle

\section{Introduction}\label{sec:Intro}

Transport in  dense liquids slows down drastically as the packing fraction is increased or the temperature is lowered.
Approaching the glass transition, transport essentially ceases since particles are trapped in long-living cages formed by their neighbors.
Then stress fluctuations require very long times to relax with the consequence that the viscosity increases drastically. The structural arrest of the liquid dynamics is accompanied by a series of characteristic features in the correlation functions, such as the emergence of stretched relaxation as terminal process and a mesoscopic time window where a critical relaxation appears. A microscopic theory for the glass transition has been achieved in terms of the mode-coupling theory~\cite{Goetze:Complex_Dynamics}, which has been tested successfully in the last 25 years.

To further elucidate the microscopic mechanism of the glass transition significant effort has been made to confine the liquid to small volumes~\cite{Alba:2006} and  thereby to identify the relevant length scale of correlated motion. Computer simulations have revealed that walls can considerably alter the transport properties~\cite{Mittal:2008} and depending on the fluid-wall interactions the confinement either facilitates the glass formation~\cite{Scheidler:2000} or suppresses the tendency towards structural arrest~\cite{Varnik:2002}.
Experiments on colloidal suspensions confined by two quasi-parallel walls
found an oscillatory dependence of the mean-square displacement perpendicular to the walls as the plate distance
is varied~\cite{Nugent:2007}. A similar variation of the glass transition line due to layering has been predicted recently by a mode-coupling theory for confined geometries~\cite{Lang:2010,Lang:thesis}.

If the liquid is confined by a random medium, for example a porous host structure, an intricate interplay of the freezing of the structural relaxation and a localization transition emerges. A series of non-trivial predictions have been made by a generalization of the mode-coupling theory accounting for the interaction with frozen obstacles~\cite{Krakoviack:2005,Krakoviack:2007,Krakoviack:2009}. The phase diagram
for mixtures of mobile and frozen components has been explored by computer simulations and semi-quantitative agreement with mode-coupling theory has been reported~\cite{Kurzidim:2009,Kurzidim:2010,Kim:2009}. If the density of the mobile component becomes very small, the interaction among the mobile particles can be ignored and the dynamics is governed solely by the excluded volume due to the frozen matrix. This limiting case is known as the Lorentz model which displays a localization transition upon crossing a critical obstacle density.
A self-consistent mode-coupling kinetic theory~\cite{Goetze:1981a,Goetze:1981b,Goetze:1982} provides a microscopic description of the phase-space correlation functions with no free parameters, which captures quantitatively the behavior for low and  moderate obstacle densities. In particular, it reproduces the known long-time anomalies in the velocity autocorrelation functions~\cite{Ernst:1971a}.
Furthermore, it correctly predicts the emergence of anomalous transport at the critical point and the emergence of scaling laws. Yet, in the immediate vicinity of the critical point deviations become manifest and it appears that mode-coupling theory in its current form is not suited to deal with divergent length scales~\cite{Goetze:Complex_Dynamics}.

In fact, the localization transition in the Lorentz model is  driven by an underlying geometric transition of the void space, i.e. the volume accessible to the mobile agents. As the obstacle density is increased, the excluded regions grow and start to merge. Then configuration space falls apart and a hierarchy of finite pockets coexists with a single spanning cluster, which  percolates through the entire system~\cite{Stauffer:Percolation}. At a certain critical density the infinite cluster ceases to have extensive weight rather it becomes a fractal. Then long-range transport occurs only on this self-similar structure leading to anomalous transport for arbitrarily long times. For even  larger densities, all clusters are finite  implying that the mean-square displacements of individual particles saturate at a finite value. In the vicinity of the critical density dynamics scaling is expected to hold~\cite{Kertesz:1983} with critical exponents from the universality class of random resistor networks~\cite{Halperin:1985,Machta:1985}.

Early computer simulations on the ballistic Lorentz model have focused on the non-analytic density dependence of the diffusion coefficient~\cite{Bruin:1972,Bruin:1974}. The long-time anomaly in the velocity autocorrelation function has been readily observed for moderate scatterer densities in two dimensions~\cite{Bruin:1972,Bruin:1974} but the prefactor appeared to be significantly different~\cite{Alder:1983,Lowe:1993} from the theoretical prediction~\cite{Ernst:1971a}. Only recently, the controversy could be resolved and has been shown that the localization transition strongly interferes with the long-time tail~\cite{Lorentz_LTT:2007}. For Brownian tracers the velocity autocorrelation function, defined via suitable derivatives of the mean-square displacements, has been calculated analytically to first order in the scattering density~\cite{Lorentz_VACF:2010} and computer simulations show  that the range of validity extends even to moderate densities~\cite{Lorentz_VACF:2010,Lorentz_2D:2010}.

The critical properties of the dynamics of the Lorentz model close to the localization transition have been tested only recently, both  in three dimensions~\cite{Lorentz_PRL:2006,Lorentz_JCP:2008,Lorentz_space:2010} as well as for the
planar case~\cite{Lorentz_2D:2010,FCS_scaling:2010}. Nice agreement with the theoretical predictions has been achieved, provided the universal, leading corrections to scaling~\cite{Percolation_EPL:2008} are taken into account. The scenario appears to be relevant also for size-disparate soft-sphere  mixtures~\cite{Moreno:2006,Voigtmann:2009,Lorentz_BSSM:2010} as well as for Yukawa mixtures~\cite{Kikuchi:2007} or fast ion diffusion in silica melts~\cite{Horbach:2002,Meyer:2004,Voigtmann:2006}.

The dynamics of the Lorentz model has been characterized so far mainly in terms of all-cluster averages, i.e. a tracer is placed at random in the void space and each pocket  contributes  to the mean with a weight given by its volume. However, long-range transport occurs only on the infinite cluster below the critical density. Here we elucidate further the dynamics of the Lorentz model, Section~\ref{Sec:model},  and study the motion of tracers that are located at the infinite cluster. The simulation details are presented in Sec.~\ref{Sec:simulation}.  Then we study the mean-square displacements for different obstacle densities (Section~\ref{Sec:msd}) and discuss the variation of the corresponding diffusivities (Sec.~\ref{Sec:diffusion}).
 The scaling behavior of the mean-square displacement
is elaborated in Section~\ref{Sec:scaling} where we also discuss the role of the leading corrections. A complementary view is to study the frequency-dependent conductivity, which is presented in Sec.~\ref{Sec:conductivity}. The transport properties beyond the second moment of the displacement is discussed in Sec.~\ref{Sec:nongaussian} exemplified for the non-gaussian parameter.

\section{Lorentz model and continuum percolation}\label{Sec:model}

A minimal model for transport in disordered media has been introduced by Lorentz~\cite{Lorentz:1905} more than a century ago. The idea is that the main contribution to electric resistance arises from the interaction of mobile agents (electrons) with frozen impurities in the medium, whereas the interaction among the mobile particles is negligible. In its simplest variant these obstacles are distributed randomly and independently in the sample and their density $n=N/V$ is sufficient to characterize the structure. The mobile agent interacts with the impurities via some potential, typically taken as hard-sphere exclusion of radius $\sigma$. Then the relevant dimensionless control parameter is the reduced obstacle density $n^* = n \sigma^3$. Since the obstacles are distributed independently, the regions of excluded volume may overlap and build large clusters as the obstacle density increases. Therefore the packing fraction $\varphi$, i.e. the volume that is inaccessible to tracers, is not merely proportional to the reduced obstacle density $n^*$. The relation between $\varphi$ and $n^*$ follows from the following simple argument. The probability that a \emph{particular obstacle} does not occupy a specified point of the large volume $V$ is just $1-4\pi \sigma^3/3 V$. Then the probability that \emph{no obstacle} covers this point is $(1 - 4\pi \sigma^3/3 V)^N$, and in  the limit of large system sizes $N\to \infty, V\to \infty$ with fixed obstacle density $n= N/V$, one obtains for the packing fraction of the excluded volume
\begin{equation}
 \varphi = 1- \exp\left( \frac{4\pi}{3} n^*\right) .
\end{equation}
Only at  low obstacle densities $\varphi  \approx 4 \pi n^*/3$ the probability of mutual overlap of the excluded volume can be ignored. For moderate $n^*$ the void space consists of a set of isolated bounded pockets coexisting with a connected region that spans through the entire system. The latter is referred to as the infinite or percolating cluster.

Above a critical obstacle density $n_c^*$ the infinite cluster ceases to exist and only the hierarchy of disconnected finite clusters prevails. The transition is of purely geometric origin and is accompanied by a series of power laws reminiscent of a continuous phase transition~\cite{Stauffer:Percolation}. Directly at the critical density $n_c^*$ the infinite cluster becomes a self-similar structure with fractal dimension\footnote{Recent values for critical exponents for percolation have been collected in Ref.~\citenum{Percolation_EPL:2008}} $d_{\rm f} = 2.53$  implying that its total weight is subextensive. The vanishing of its weight $P_\infty$
as the  critical density is approached from below, $n^*\uparrow n_c^*$, is again determined  by a power law $P_\infty \sim (-\epsilon)^\beta$, where $\epsilon = (n^*-n_c^*)/ n_c^*$ denotes the separation parameter, a convenient measure for the distance to the critical point. The critical exponent $\beta$ assumes the value $\beta = 0.41$
for three-dimensional percolation. The infinite cluster remains fractal below a divergent length scale $\xi \sim |\epsilon|^{-\nu}$ known as the correlation length and appears homogeneous only on larger scales. A simple scaling argument reveals that the new critical exponent $\nu$  is connected to  the two introduced above by a hyperscaling relation  $d_{\rm f} = d- \beta/\nu$ and in the three-dimensional case one obtains $\nu = 0.88$. The structural properties of the infinite cluster and the coexisting hierarchy of finite clusters in the immediate vicinity of the geometric transition is then characterized by two independent universal critical exponents. By universality,  introducing short-range correlations in the distribution of the obstacles is expected not to alter the nature of the percolation and to leave the values of the exponents unchanged. Similarly the scaling functions are considered to be universal and details should only enter  the amplitudes, e.g. how the correlation length is connected to the radius $\sigma$ of the obstacles.

The dynamics of particles on such fractal objects becomes self-similar again and introduces new critical exponents, which are independent of the ones discussed above. Here,  we consider a ballistic motion for the tracer  scattering elastically at the frozen obstacles. Then the magnitude of the velocity $v = |\vec{v}| $ is conserved and only the direction of $\vec{v}$ is changed each time a scattering event occurs. For a particle initially located on the infinite cluster the linear extension $R$ of the probability cloud  to find the particle at some distance  after an elapsed time $t$ is expected to grow as a power law $R \sim t^{1/d_{\rm w}}$ directly at the percolation transition. Here the walk dimension $d_{\rm w}$ generalizes the concept of Brownian motion where $ R \sim t^{1/2}$ also leads to a self-similar increase. These arguments suggest that the mean-square displacement of the tracer $\delta r^2_\infty(t)$ becomes subdiffusive $\delta r^2_\infty(t) \sim t^{2/d_{\rm w}}$ for long times  never reaching the regime of diffusion. The mechanism for this phenomenon is encoded in the fractal structure of the configuration space and naturally explains the emergence of anomalous transport in complex structures.

The transport properties  in the Lorentz model close to the transition point again constitute  a critical phenomenon with a universality class that is shared by certain random resistor networks. There, the fundamental quantity of interest is the vanishing of the conductivity, e.g. by cutting randomly bonds of a regular lattice of nodes connected by resistors to their nearest neighbors. Then at the percolation transition the residual network becomes a fractal with zero linear conductivity. Since the geometric properties of lattice percolation and
the continuum percolation underlying the localization transition of the Lorentz model have been shown to be
identical~\cite{Elam:1984}, it is suggestive that also the dynamic critical  properties are related. There is an interesting subtlety though that requires further insight. In continuum percolation the obstacles can come arbitrarily close to each other, leading to narrow channels through which particles have to squeeze through. These bottlenecks arising due to purely geometric reasons weakly connect different compartments of a cluster and the distribution of exit probabilities from such compartments is again power-law distributed. Then the mapping of the Lorentz model to a random resistor network on a lattice requires to consider power-law distributed
conductances~\cite{Halperin:1985,Machta:1985} between nearest neighbors that are diluted until the percolation transition is reached.

\section{Simulation details}\label{Sec:simulation}

Our computer simulations are event-driven molecular dynamics simulations of a single tracer particle in a background of randomly distributed, overlapping hard spheres.
We use a cubic container of linear box size $L=100 \sigma$ with periodic boundaries, thus our systems contain 300,000 to 840,000 obstacles. To study the influence of the finite system size, a number of systems with $L=50\sigma$ and $L=150\sigma$ were also tested. The remaining infinite cluster of void space is identified by calculating the Voronoi diagram~\footnote{We have employed the free voro++ package initially developed by Chris Rycroft, see \url{http://math.lbl.gov/voro++/}.} of all obstacle centers and analyzing the network of edges not overlapping with the spheres.
A cluster is regarded as infinite if there is an infinite path for a point-like tracer, i.e. the network wraps around of at least one of the boundaries~\cite{Spanner:thesis}.

For most densities, 16 different obstacle realizations were analyzed with 5 trajectories each starting at different points, making a total of 80 independent trajectories per density. At $n^*=0.83$ the number of systems was increased to 80, of which 76 contained an infinite cluster, and at $n^*=0.84$, there was an infinite cluster in 27 out of 100 tested systems.
While the calculation of trajectories is purely limited by CPU time (one trajectory took about 3 hours on a 3.0 GHz Intel Xeon 5160, independently of the system size), the amount of available memory limits the number of obstacles for the analysis of the cluster structure (a system of 1,000,000 obstacles required approximately 3 GB of RAM).

\section{Mean-square displacement}\label{Sec:msd}
To characterize the motion of the tracer on the infinite cluster, we have first calculated the mean-square displacements $\delta r^2_\infty(t) = \langle [ \vec{R}(t) - \vec{R}(0)]^2 \rangle_\infty$ for obstacle densities up to the percolation transition. Here $\vec{R}(t)$ denotes the position of the tracer and brackets $\langle \cdot \rangle_\infty$ denote a moving time average, an average over different trajectories, as well as the different realizations of the disorder. The subscript $\infty$ indicates that the particle starts and remains on the infinite cluster.

\begin{figure}
\centering
\includegraphics[width=\figurewidth]{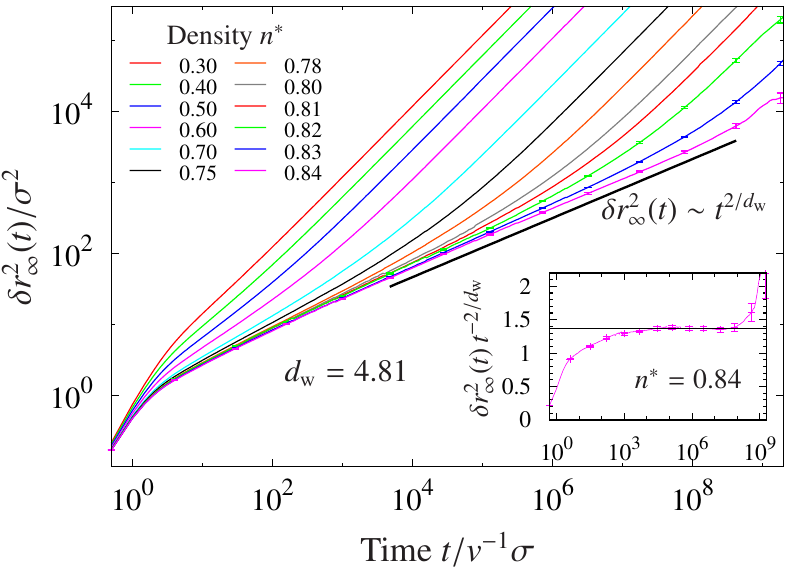}
\caption{Mean-squared displacements for ballistic tracers on the infinite cluster. Obstacle density increases from top to bottom. The straight line corresponds to a power law $t^{2/d_w}$ and serves as a guide to the eye.  }
\label{fig:msd_infinite}
\end{figure}

The results for $\delta r^2_\infty(t)$ are displayed in Fig.~\ref{fig:msd_infinite} on double logarithmic scales. The mean-square displacements extend over more than eight orders of magnitude in time and span length scales of a factor of 150 particle diameters thereby displaying essentially no noise. For short times the data for all densities collapse on a single straight line which reflects the free ballistic motion, $\delta r^2_\infty(t) = v^2 t^2$. The mean free path length before the particle encounters an obstacle can be estimated by $1/ n  \sigma^2$ implying a collision rate $n v \sigma$. Since here we consider moderate densities, we measure times in units of $t_0 := \sigma/v$ which then is  comparable to the time of free flight until the next collision event. For rather low scatterer density the mean-square displacements cross over directly   to a diffusive behavior $\delta r^2_\infty(t) = 6 D_\infty t$. Increasing  the obstacle density, the long-time behavior gradually shifts to the right, implying that the diffusion coefficient $D_\infty$ is suppressed.

At a critical density $n_c^* = 0.839$ the diffusive regime is never reached in the window of the simulation time. Rather,  the mean-square displacement increases subdiffusively, $\delta r^2_\infty(t) \sim t^{2/d_{\rm w}}$, over more than six decades in time. The exponent nicely corroborates the value $d_{\rm w} = 4.81$ which is expected from scaling predictions, compare e.g.~\cite{Percolation_EPL:2008,benAvraham:DiffusionInFractals}. The critical density $n_c^*$ also coincides with the point  above which no percolating cluster exists for large system sizes.

Just below the critical density the mean-square displacements follow the critical law up to a certain time scale and a crossover to  diffusive behavior is observed. As the critical density is approached the crossover shifts to later times or equivalently to larger distances. The shape of the crossover approaches a universal form and scaling behavior is expected, which we analyze in Sec.~\ref{Sec:scaling}.

\section{Diffusion coefficients}\label{Sec:diffusion}

\begin{figure}
\centering
\includegraphics[width=\figurewidth]{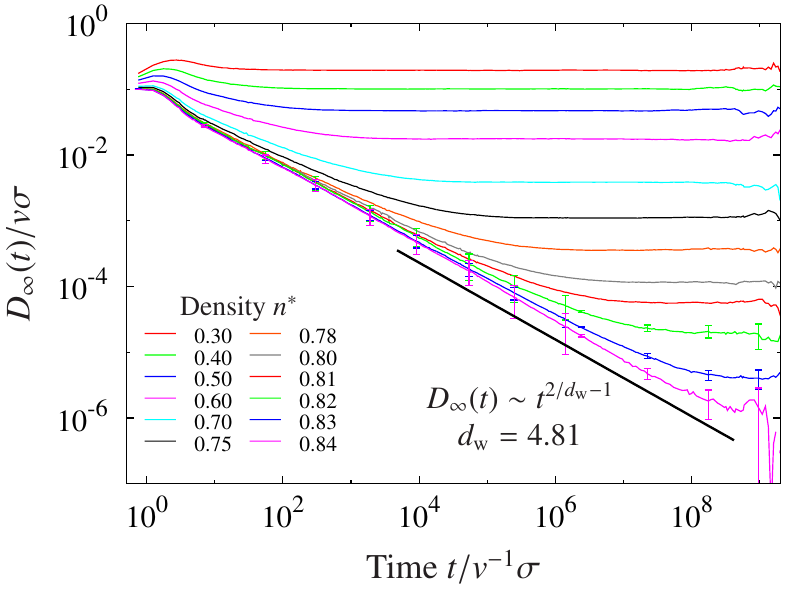}
\caption{Time-dependent diffusion coefficient $D_\infty(t) = (1/6) \diff \delta r^2_\infty(t)/\diff t$ for particles confined to the infinite cluster. Obstacle density increases from top to bottom. At the localization threshold a power law $t^{2/d_{\rm w}-1}$ becomes manifest, indicated by a straight line as guide to the eye.}
\label{fig:time_dependent_diffusion3d_infinite}
\end{figure}

A different representation of the transport properties encoded in the  mean-square displacements is given in terms of the time-dependent diffusion coefficient
\begin{equation}\label{eq:time_diffusion}
 D_\infty(t) = \frac{1}{2d} \frac{\diff}{\diff t } \delta r^2_\infty(t) \, ,
\end{equation}
where $d=3$ is the dimension of the surrounding space. Figure~\ref{fig:time_dependent_diffusion3d_infinite} displays $D_\infty(t)$ obtained by numerical derivatives of the mean-square displacements. Due to the blocking algorithm the data are stored essentially on a logarithmic grid such that  the derivatives do not introduce significant  new noise. The short-time ballistic motion becomes manifest as linear increase of $D_\infty(t)$ (not shown) before a maximum value is reached. At longer times  the time-dependent diffusion coefficient drops again, implying that the obstacles induce anticorrelations in the velocity autocorrelations. For long times $D_\infty(t)$ saturates and assumes the long-time diffusion coefficient $D_\infty := D_\infty(t\to\infty)$. The time-dependent diffusion coefficient, defined in Eq.~(\ref{eq:time_diffusion}),
converges much more rapidly to its long-time limit than $\delta r^2_\infty(t) / 6t  $, since it only probes properties at late times. In our simulations, the diffusion coefficient drops by five orders of magnitude as the critical obstacle density is approached. Then the subdiffusive motion becomes apparent as a power-law decay of $D_\infty(t)$.

\begin{figure}
\centering
\includegraphics[width=\figurewidth]{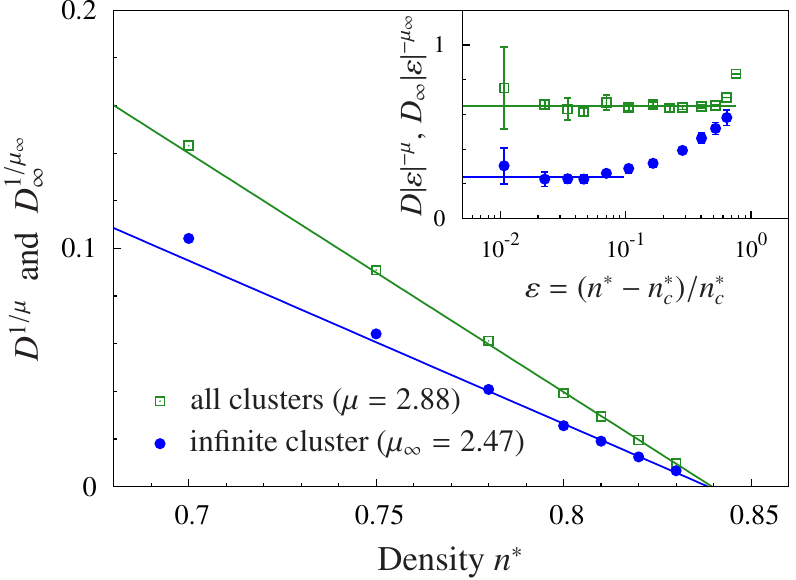}
\caption{Rectification of the diffusion coefficients for the infinite cluster $D_\infty$ and the all-cluster average $D$. Both diffusion coefficients vanish at the same critical obstacle density $n_c^* =0.839$. Inset: Scaling behavior of the diffusion coefficients with the separation parameter.}
\label{fig:diffusion_infinite}
\end{figure}

The extracted diffusion coefficients are expected to vanish as a power law $D_\infty(n) \sim (n_c^* - n^*)^{\mu_\infty}$ with a suitable critical exponent $\mu_\infty$. Recently, it was shown that the diffusion coefficient for an all-cluster average vanishes as $D(n)
\sim (n_c^* - n^*)^{\mu}$ with a value $\mu = 2.88$~\cite{Lorentz_PRL:2006,Lorentz_JCP:2008}. Since only particles that are located in the percolating clusters can contribute to diffusion, we anticipate $D(n) \sim D_\infty(n) P_\infty(n)$ and infer the scaling relation
$\mu_\infty = {\mu} - \beta = 2.47$. The prediction can be tested in the most sensitive way by rectification plots, see Fig.~\ref{fig:diffusion_infinite}. There both $D(n)^{1/\mu}$ and $D_\infty(n)^{1/\mu_\infty}$ become straight lines as a function of the scatterer density $n^*$, at least for densities above  $n^* = 0.78$. Furthermore, they extroplate to the same critical density $n_c^* =0.839$ which coincides with the density where no cluster percolates through the entire system. The scaling behavior is demonstrated also in the inset of Fig.~\ref{fig:diffusion_infinite} where we show that $D(n) | \epsilon |^{-\mu}$ and  $D_\infty(n) |\epsilon|^{-\mu_\infty}$ approach constants as the separation parameter $\epsilon$ becomes smaller.

\section{Scaling behavior}\label{Sec:scaling}
The similarity of the mean-square displacements and the power-law dependence of the diffusion coefficient in the vicinity of the percolation transition suggests a scaling law. Then $\delta r_\infty^2(t;n)$ is no longer a function of both time and density but the density dependence can be hidden by suitable shifts in time and length scales. Directly at $n_c^*$ the critical law $\delta r_\infty^2 \sim t^{2/d_{\rm w}}$
persists for all times suggesting the scaling hypothesis
\begin{equation}\label{eq:scaling_hypothesis}
 \delta r^2_\infty(t;\epsilon) = t^{2/d_{\rm w}} \delta \hat{r}_\infty^2(t/t_x),
\end{equation}
where the time scale $t_x$ diverges for $n^* \uparrow n^*_c$.  For short rescaled times, the scaling function assumes a constant $\delta \hat{r}^2_\infty( \hat{t} \ll 1)  = const.$, since the particle has not explored the infinite cluster on large enough scales to sense deviations from a fractal structure. On the other hand, for large rescaled time the mean-square displacement increases linearly implying $\delta \hat{r}^2_\infty(\hat{t} \gg 1) \sim \hat{t}^{1- 2/d_{\rm w}}$. Furthermore, one expects that the crossover from subdiffusive to linear increase of the mean-square displacement $\delta r^2_\infty(t)$ occurs on a length scale given by the correlation length $\xi$, $\delta r^2_\infty(t_x) \sim \xi^2$, suggesting a power-law  divergent crossover time scale $t_x \sim \xi^{d_{\rm w}} \sim (-\epsilon)^{-\nu d_{\rm w}}$.  For the long-time diffusion coefficient one infers $D_\infty \sim t_x^{2/d_{\rm w}-1} \sim \xi^{2-d_{\rm w}}$ and we obtain a relation between the critical exponents
$\mu_\infty = \nu (d_{\rm w}-2)$~\cite{benAvraham:DiffusionInFractals}. The values that we have extracted from the subdiffusive increase of the mean-square displacement $\delta r^2_\infty(t)$ and the vanishing of the corresponding diffusion coefficient $D_\infty$ nicely corroborate the exponent relation.

Earlier it has been demonstrated that the all-cluster averaged mean-square displacement also displays subdiffusive behavior $\delta r^2(t) \sim t^{2/z}$ at the critical obstacle density~\cite{Lorentz_PRL:2006,Lorentz_JCP:2008}, yet with a different dynamic exponent $z=6.25$. Here the self-similar hierarchy of finite clusters contributes to the mean-square displacement. Below the critical density $n^* < n_c^*$ a regime is reached where $\delta r^2(t)$ increases again linearly, $\delta r^2(t) = 6 Dt$, and $D \sim (-\epsilon)^{\mu}, \mu = \mu_\infty + \beta$ decreases faster than $D_\infty \sim (-\epsilon)^{\mu_\infty}$ since only the particles initially on the infinite cluster contribute to long-range transport. The time needed for these  particles to explore the infinite cluster is still given by $t_x \sim \xi^{d_{\rm w}}$ and
and the scaling law
$ \delta r^2(t;\epsilon) = t^{2/z} \delta \hat{r}^2(t/t_x)$
is anticipated. The scaling function $\delta \hat{r}(\hat{t})$ encodes the crossover from anomalous transport $\delta \hat{r}^2(\hat{t} \ll 1) = const.$ to the linear regime $\delta \hat{r}^2(\hat{t} \gg 1) \sim \hat{t}^{1-2/z}$. In particular, the scaling hypothesis predicts that the diffusion coefficient vanishes as $D \sim t_x^{2/z-1} \sim (-\epsilon)^\mu$ which leads to the exponent relation
$ \mu = \nu d_{\rm w} (1-2/z)$.  Combining with the scaling relation obtained so far, one finds for the dynamic exponent $z = \nu d_{\rm w}/(\nu - \beta/2)$ and the measured value $z=6.25$ is in excellent agreement with the prediction.
Second, the length scale $\ell$ where the crossover from subdiffusive to linear increase occurs, $\ell^2 \sim \delta r^2(t_x)$
is given by $\ell \sim t_x^{1/z}$. Employing the scaling relations derived so far, one finds $\ell \sim (-\epsilon)^{-\nu + \beta/2}$ and it is determined entirely by critical exponents for structural properties of the percolation transition. In particular it diverges slower than the correlation length $\xi \sim (-\epsilon)^{-\nu}$ and in fact constitutes the second moment of the size distribution of the clusters~\cite{Stauffer:Percolation} and the measured values of $\ell$ were shown to corroborate the scaling hypothesis~\cite{Lorentz_PRL:2006,Lorentz_JCP:2008}.

\begin{figure*}
\centering
\includegraphics[width=\linewidth]{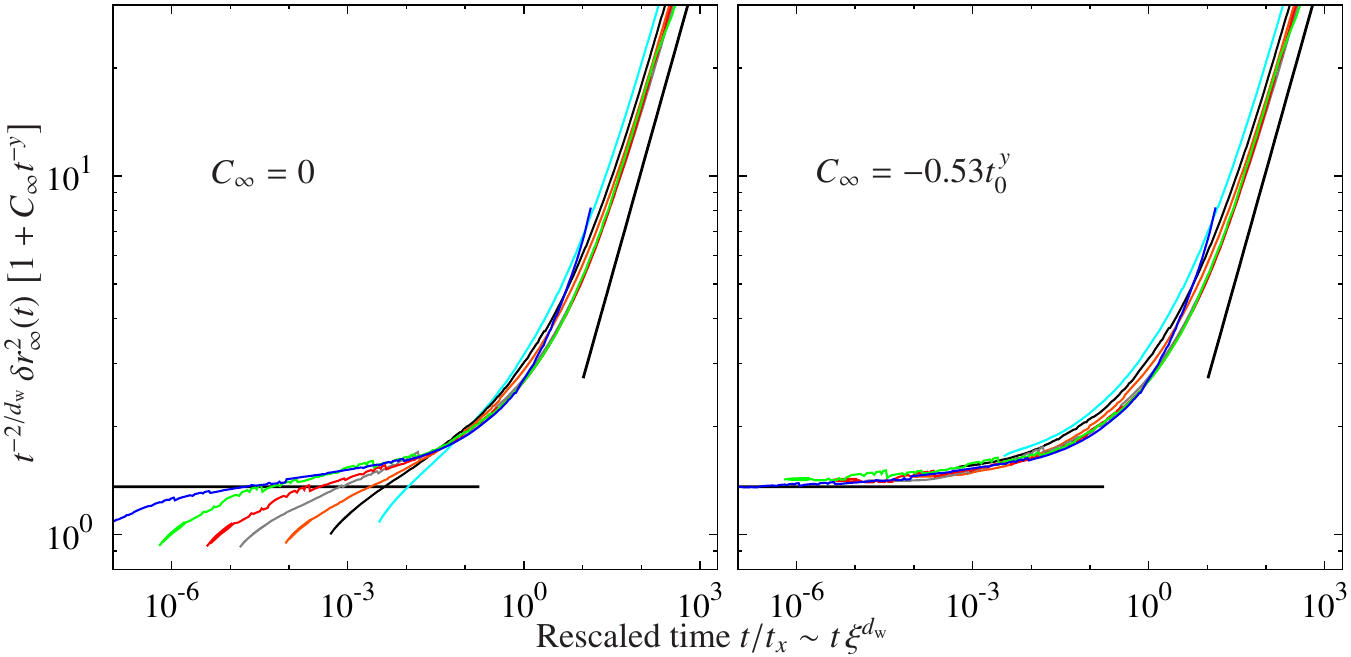}
\caption{Scaling behavior of the mean-squared displacement on the infinite cluster. Left panel: Upon rescaling with the crossover time $t_x := t_0(-\epsilon)^{-\nu d_{\rm w}}$ with $t_0 ;= \sigma/v$ the data converge to a master curve as the critical density is approached.
Right panel: Leading corrections are approximately taken into account by adjusting a single nonuniversal amplitude $C_\infty = -0.53 t_0^y$. }
\label{fig:msd_infinite_scaling}
\end{figure*}

The consistency of the values of the exponents expected from the exponent relations is encouraging to test the scaling hypothesis, Eq.~(\ref{eq:scaling_hypothesis}) itself. Thus upon dividing by the critical law $t^{2/d_{\rm w}}$ and measuring times in units of the crossover time $t_x \sim \xi^{d_{\rm w}}$ all data sets are expected to collapse onto a single master curve $\delta \hat{r}_\infty^2(\hat{t})$, see Fig.~\ref{fig:msd_infinite_scaling}.
The curves all display the same long-time asymptote and exhibit a satisfactory data collapse. However, this reflects only that for all data sets our simulations reach the diffusive regime and that the diffusion coefficients follow the scaling prediction. Deviations from scaling become manifest at low rescaled times and here one infers only that the data sets approach a master curve as the separation parameter to the critical point is decreased. The crossing of the asymptote occurs where the scaling regime merges continuously with the microscopic time scales. The fanning out of the data for short rescaled times has also been observed for the all-cluster averaged mean-square displacements~\cite{Lorentz_PRL:2006,Lorentz_JCP:2008} and it was argued that corrections to scaling should be taken into account.

The idea in the renormalization group is that  rescaling of length scales corresponds to changing the coupling constants in some large dimensional space of parameters. The critical point corresponds to a fixed point of the renormalization group flow and the properties of this fixed point determine the values of the critical exponents. The scaling functions encode how the critical point repels trajectories that are close to but not on  the critical manifold. In particular, the corrections to scaling are universal again and contain the information how the
scaling regime is approached or how microscopic details become irrelevant. These ideas suggest that the scaling hypothesis, Eq.~\ref{eq:scaling_hypothesis} can be extended to account for generic correction terms~\cite{Percolation_EPL:2008}
\begin{equation}
 \delta r^2(t;\epsilon) = t^{2/d_{\rm w}} \delta \hat{r}^2_\infty(\hat{t}) \left[ 1 + t^{-y} \Delta_\infty(\hat{t}) \right], \quad \hat{t} = t/t_x.
\end{equation}
The exponent $y$ is again universal and has been determined to $y=0.34$ for the three-dimensional Lorentz model from a similar scaling hypothesis for the all-cluster averaged mean-square displacement~\cite{Lorentz_PRL:2006,Lorentz_JCP:2008}. The new scaling function assumes a constant for small rescaled times $C_\infty := \Delta_\infty(\hat{t} \ll 1)$ and behaves like a power-law at large scaling times
$\Delta_\infty(\hat{t} \gg 1) = C_D^{(\infty)} \hat{t}^y$. In principle, the amplitude $C_\infty$ can be inferred from the mean-square displacement at the critical obstacle density
\begin{equation}
 \delta r^2(t;\epsilon=0) \sim t^{2/d_{\rm w}} \left[ 1 + C_\infty t^{-y} \right] \, .
\end{equation}
Similarly, the amplitude $C_D^{(\infty)}$ encodes the corrections to scaling for the diffusion coefficient
\begin{equation}
 D_\infty(\epsilon) \sim (-\epsilon)^{\mu_\infty} \left[ 1 + C_D^{(\infty)} t_0^{-y} (-\epsilon)^{- y \nu d_{\rm w}} \right]
\end{equation}
Since the detailed shape of $\Delta_\infty(\hat{t})$ is unknown, one has to rely on simple approximation schemes. A convenient interpolation function could be $\Delta_\infty(\hat{t}) = C_\infty + C_D^{(\infty)} \hat{t}^y$. Since in our data the corrections to scaling are most significant at short rescaled times, we ignore the corrections to the diffusion coefficient $C_D^{(\infty)} \approx 0$,
and approximate $\Delta_\infty(\hat{t}) = C_\infty$ for all times $\hat{t}$. The remaining unknown coefficient $C_\infty$ is adjusted to yield an optimal data collapse, see Fig.~\ref{fig:msd_infinite_scaling}. Choosing $C_\infty = -0.53\, t_0^y$ the data  nicely collapse on a single master curve although the improvement is less satisfactory than for the all-cluster average~\cite{Lorentz_PRL:2006,Lorentz_JCP:2008} or in two-dimensional transport
on a lattice~\cite{Percolation_EPL:2008}. The reason may be traced back to the slower convergence of the diffusion coefficients
identified in the inset of Fig.~\ref{fig:diffusion_infinite}.

\section{Conductivity}\label{Sec:conductivity}

In applications, such as ion transport in disordered solids, the basic quantity of interest is the frequency-dependent complex conductivity $\sigma(\omega)$. Applying a small  homogeneous alternating electric field $\vec{E}(t) = {\rm Re}[ \vec{E}(\omega) {\rm e}^{- {\rm i} \omega t}] $ to the sample the resulting electric current density will alternate with the same frequency $\vec{j}(t) = {\rm Re}[ \vec{j}(\omega)  {\rm e}^{- {\rm i} \omega t}] $ in the linear response and the appropriate frequency-dependent generalization of Ohm's law reads $\vec{j}(\omega) = \sigma(\omega) \vec{E}(\omega)$. The real part  $\sigma'(\omega) = {\rm Re}[ \sigma(\omega)]\geq 0$ describes the dissipation due to friction, whereas the  imaginary part $\sigma''(\omega) = {\rm Im}[\sigma(\omega)]$ is connected to the energy stored in the system.

The current density constitutes a collective property of the sample but for dilute charge carrier concentration $n_{\rm ion}$ and charge density $q n_{\rm ion}$ the interaction between the mobile charges may be ignored. Then  each ion experiences a frequency-dependent  force $\vec{F}(\omega) = q \vec{E}(\omega)$ resulting in an oscillatory velocity $\vec{v}(\omega) = \mu(\omega) \vec{F}(\omega)$, where $\mu(\omega)$ denotes the frequency dependent mobility, sometimes also referred to as admittance~\cite{Felderhof:2005,Jeney:2008,Franosch:2009}. Since for independent ions the total current density is given by $\vec{j}(\omega) = n_{\rm ion} q \vec{v}(\omega)$
the frequency-dependent conductivity can be related to the frequency-dependent mobility $\mu(\omega)$ by $\sigma(\omega) =  q^2 n_{\rm ion}   \mu(\omega)$. By the well-known fluctuation-dissipation theorem~\cite{Kubo:Statistical_Physics_II}
$\mu(\omega) =  Z(\omega)/k_B T$ where
\begin{equation}
 Z(\omega) = \frac{1}{d} \int_0^\infty \diff t \, {\rm e}^{ {\rm i} \omega t} \langle \vec{v}(t) \cdot \vec{v}(0) \rangle ,
\end{equation}
is the one-sided Fourier transform of the velocity autocorrelation function. In particular,  the Green-Kubo relation is obtained as low-frequency limit $D= Z(\omega\to 0)$ and one recovers the Stokes-Einstein-Sutherland relation $\mu = D/k_B T$.
In summary, $Z(\omega)$ encodes the frequency-dependent response to an alternating field and coincides up to well-understood factors with the frequency-dependent conductivity.

Here we single out the conductivity for the particles located on the infinite cluster only. The experimental condition in mind is that the system initially does not contain ions but they are inserted in the infinite cluster by wiring the surfaces. Then the measured conductivity
is essentially $Z_\infty(\omega)$, viz. the one-side Fourier transform of the velocity autocorrelation, where the average is restricted to particles on the spanning cluster.

\begin{figure}
\centering
\includegraphics[width=\figurewidth]{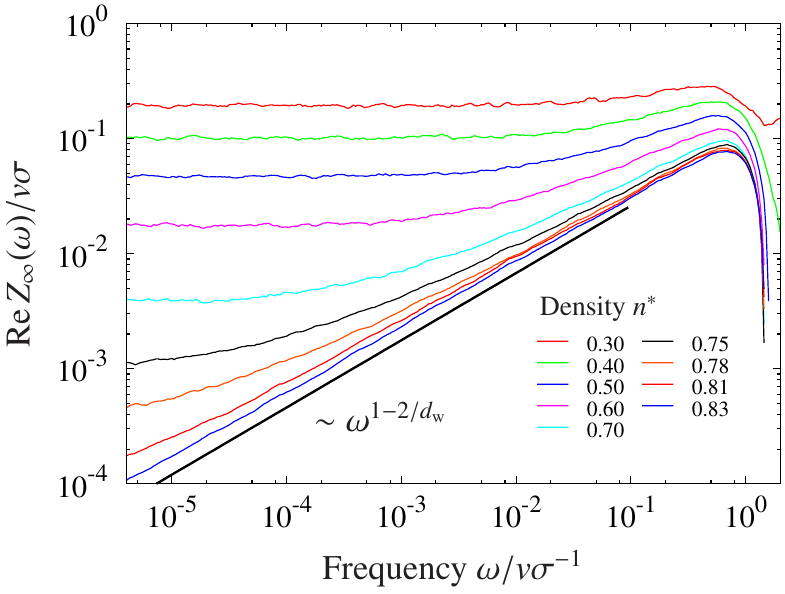}
\caption{Real part of the frequency-dependent conductivity for particles restricted to the infinite cluster. Obstacle density increases from top to bottom. At the critical obstacle density $n_c^*$ the conductivity follows a power law indicated by the black line as guide to the eye. }
\label{fig:frequency-diffusion}
\end{figure}

The numerical evaluation of $Z_\infty(\omega)$  is achieved upon integrating by parts
\begin{equation}
 Z_\infty(\omega) = D_\infty - {\rm i} \omega \int_0^\infty [ D_\infty(t) - D_\infty ] {\rm e}^{ {\rm i} \omega t} \diff t ,
\end{equation}
and the Fourier transform can be easily performed relying on the simplified Filon algorithm~\cite{Tuck:1967}. The loss part of the conductivity ${\rm Re}[ Z_\infty(\omega)]$ is displayed in Fig.~\ref{fig:frequency-diffusion} for a frequency window covering more than five decades and three orders of magnitude in amplitude. At the critical density $n_c^*$,  anomalous transport becomes manifest in terms of a  power-law $\omega^{1-2/d_{\rm w}}$. This fractal behavior is inherited directly from the power-law decay of the time-dependent diffusion coefficient $D_\infty(t) \sim t^{2/d_{\rm w} -1}$ or equivalently by the subdiffusive increase of $\delta r^2_\infty(t) \sim t^{2/d_{\rm w}}$.

For lower scatterer densities $n^* < n_c^*$ the curves display a finite low-frequency plateau which decreases drastically as the percolation threshold is approached. One identifies this low-frequency limit with the diffusion coefficient as $D_\infty = Z_\infty(\omega)$ or with the {\it dc} conductivity, respectively. In the vicinity of $n_c^*$ the curves approach the power law increase $\omega^{1-2/d_{\rm w}}$ in an intermediate time window that gradually opens as $n^* \uparrow n_c^*$. The crossover from the {\it dc} conductivity plateau to the critical behavior occurs at a crossover frequency $\omega_x$ that shifts to lower frequencies as the density comes closer to the percolation transition. Upon inspection of the curves one anticipates scaling behavior
\begin{equation}
 Z_\infty(\omega;\epsilon) = \omega^{1-2/d_{\rm w}} {\cal Z}_\infty(\hat{\omega}) , \qquad \hat{\omega} = \omega/\omega_x ,
\end{equation}
where the scaling function ${\cal Z}_\infty(\hat{\omega})$ encodes the crossover. By construction ${\cal Z}_\infty(\hat{\omega} \gg 1) = const.$, whereas the behavior for low rescaled frequencies follows from the requirement that the conductivity remains finite, ${\cal Z}_\infty(\hat{\omega}) \sim \hat{\omega}^{2/d_{\rm w}-1}$. Consequently the diffusion coefficient scales as $D_\infty \sim \omega_x^{1-2/d_{\rm w}}$ and from the previously derived scaling relations one infers $\omega_x \sim (-\epsilon)^{\nu d_{\rm w}}$. In particular, the crossover frequency of the conductivity scales as the inverse of the crossover frequency in the mean-square displacement $\omega_x \sim t_x^{-1}$ as anticipated from the Fourier transforms.

\section{Non-gaussian parameter}\label{Sec:nongaussian}
For ordinary diffusion, the statistics of the increments $\Delta \vec{R}(t) := \vec{R}(t)-\vec{R}(0)$ is given by a gaussian distribution, characterized entirely by the first two cumulants. The first vanishes by rotational symmetry  $\langle \Delta \vec{R}(t) \rangle = 0$, whereas the second is simply the mean-square displacement $\delta r^2(t) = \langle [\Delta \vec{R}(t)]^2 \rangle = 6 D t$. The basic assumption underlying the picture of ordinary diffusion is that the increments are independent at least for sufficiently long  lag time  $t$ and the central limit theorem applies. Then by independence $\delta r^2(t) \sim t$ increases linearly and the diffusion coefficient $D$ characterizes the variance of the increments for a  unit time.

\begin{figure}
\centering
\includegraphics[width=\figurewidth]{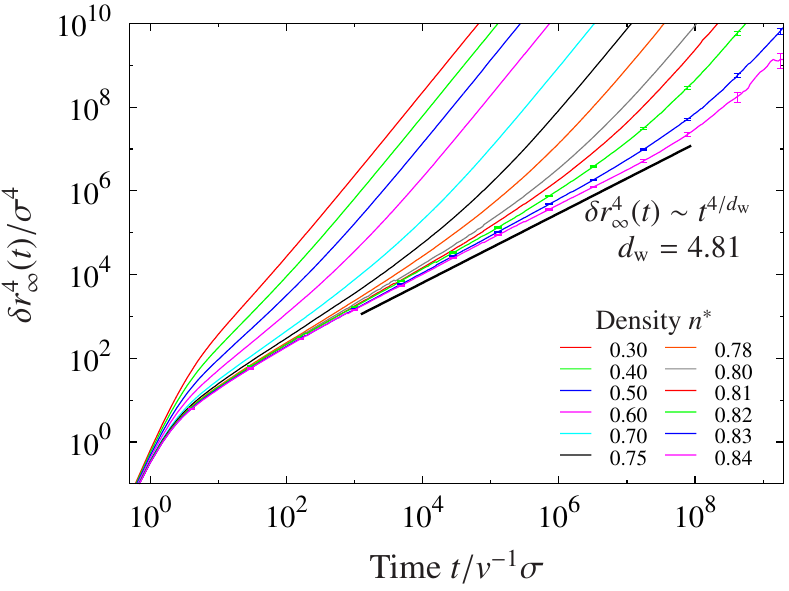}
\caption{Mean quartic displacement $\delta r^4_\infty(t)$ for particles restricted initially to the infinite cluster.
Obstacle density increases from top to bottom. At the localization transition $\delta r^4_\infty$ increases as a power-law $t^{4/d_{\rm w}}$ which is indicated by a straight line as guide to the eye. }
\label{fig:mqd_infinite}
\end{figure}

In the case of subdiffusion the central limit theorem is necessarily violated, and the higher cumulants encode valuable information on the statistics of the increments.
By rotational symmetry all odd moments still vanish identically, and here we focus on the next higher nontrivial moment $\delta r_\infty^4(t) := \langle | \Delta\vec{R}(t)|^4 \rangle_\infty$, where we restrict again the average to particles confined to the infinite cluster. The simulation results are displayed in Fig.~\ref{fig:mqd_infinite}
and one identifies a power-law increase at the critical density $\delta r_\infty^4(t) \sim t^{4/d_{\rm w}}$ over more than six decades in time. The data for very long times are presumably affected by statistical fluctuations, viz. whether an infinite cluster is present or not. Since we have averaged the data over those realizations that in fact display such an infinite cluster,  the curves are expected to display some statistical bias.
 For densities below the percolation threshold the data  reach  long-time asymptotes $\delta r_\infty^4(t) \sim t^2$ as expected for diffusive motion, and the crossover time is identified with the one of the mean-square displacements (not shown).

\begin{figure}
\centering
\includegraphics[width=\figurewidth]{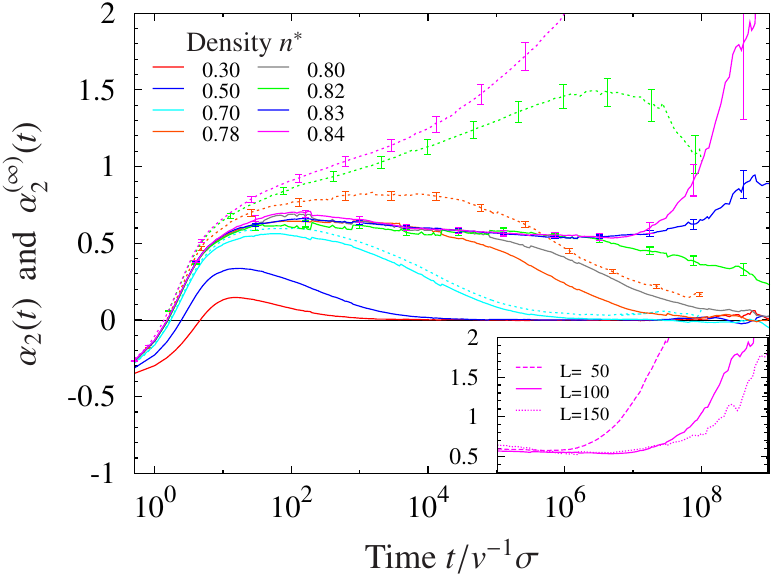}
\caption{Non-gaussian parameter for particles confined to the percolating cluster. The full lines correspond to particles restricted to the percolating cluster, whereas the broken lines are the all-cluster-averaged dynamics.
}
\label{fig:ngp_infinite}
\end{figure}

A more sensitive quantity characterizing  the differences to a gaussian statistics of the increments is the dimensionless non-gaussian parameter
\begin{equation}
 \alpha_2^{(\infty)}(t) = \frac{3}{5} \frac{\delta r_\infty^4(t)}{[\delta r_\infty^2(t)]^2} -1 ,
\end{equation}
where the prefactors are chosen such that for a three-dimensional gaussian transport $\alpha_2^{(\infty)}(t) \equiv 0$ for all times. The non-gaussian parameter has been identified in the context of supercooled liquids and dense colloidal suspensions as an indicator for heterogeneous dynamics, i.e. a growth of the non-gaussian parameter in
an intermediate time window suggests that several processes with a broad spectrum of individual relaxation times are present. In Fig.~\ref{fig:ngp_infinite} we display the non-gaussian parameter for particles confined to the infinite cluster as well as the one corresponding to the all-cluster-averaged motion $\alpha_2(t) = (3/5) \delta r^4(t)/ [\delta r^2(t)]^2-1$. The most prominent difference is that $\alpha_2^{(\infty)}(t)$ approaches zero for long times for densities below the localization threshold, whereas $\alpha_2(t\to \infty)$ saturates at finite values for $n^* < n^*_c$, which grow upon approaching the transition. Directly at the critical point $\alpha_2(t)$ is predicted  to diverge as a power law~\cite{Lorentz_PRL:2006}, in contrast to the finite limiting value $\alpha_2^{(\infty)}(t\to \infty)$. The data for $\alpha_2^{(\infty)}(t)$ at $n_c^*$ are sensitive to simulation details and  appear to increase drastically at late times. The inset in Fig.~\ref{fig:ngp_infinite} tests the long-time behavior of $\alpha_2^{(\infty)}(t)$ for different box sizes and we conclude that the observed increase is indeed a finite-size effect.

The decay to zero at late times of $\alpha_2^{(\infty)}(t)$ for densities below $n_c^*$ suggests that the statistics of increments, once the dynamics is restricted to the percolating cluster, becomes a gaussian. To characterize the motion on the infinite cluster beyond the lowest moments one should calculate the van-Hove correlation function $G_\infty(r,t) := \langle \delta[\vec{r} - \Delta\vec{R}(t)] \rangle_\infty$, which represents the probability for the tracer to have moved a distance $\vec{r}$ in a lag time $t$ on the infinite cluster. In particular, all moments of the distribution of $\Delta \vec{R}(t)$ can be calculated. Scaling suggests that at criticality
\begin{equation}
 G_\infty(r,t) = r^{-d} \widehat{G}_\infty( \hat{t} ) \, , \qquad \hat{t} = t/t_x \sim t \xi^{-d_{\rm w}}
\end{equation}
implying that the even  moments scale as $\delta r^{2n}(t) \sim t^{2n/d_{\rm w}}$~\cite{benAvraham:DiffusionInFractals} with nontrivial amplitudes determined from the scaling function $\widehat{G}_\infty(\cdot)$. Below $n_c^*$ the infinite cluster appears homogeneous on length scales larger than the correlation length $\xi$ and one expects to recover the gaussian propagator for the motion.

\enlargethispage{\baselineskip}

Let us emphasize again that the behavior of the all-cluster-averaged motion is qualitatively different from the infinite-cluster-only dynamics. The latter displays anomalous transport with propagators that deviate from simple gaussians for length scales  $r \ll \xi$ and time scales $t \ll t_x$, whereas the long-time large-distance motion is described again by simple diffusion as anticipated by the central limit theorem. In contrast, the all-cluster-averaged motion  never reaches the diffusion law,
although the corresponding mean-square displacement increases linearly for times larger than the crossover time $t_x$. The frozen disorder implies that configuration space is compartmentalized for all densities such that the particles confined to some finite cluster remain there for all times. Thus the all-cluster-averaged dynamics is intrinsically heterogeneous at all densities and all times.

\section{Summary and Conclusion}\label{Sec:summary}

Long-range transport in compartementalized solids is restricted to the infinite cluster percolating through the entire system. We have established within the Lorentz model as minimal model for such heterogeneous media that the arrest of the dynamics coincides with the percolation threshold. Then the infinite cluster becomes a fractal object and transport on it  becomes anomalous. Here we have employed large scale computer simulations to test the dynamics at the localization threshold and in its immediate vicinity. We have corroborated that the dynamics can be rationalized by a critical phenomenon, similar to a continuous phase transition, and tested various aspects of the critical behavior. In particular, we find that three independent critical exponents are necessary to obtain a complete description of the dynamics. Here the exponent $\beta$ quantifies the vanishing of  the weight of the infinite cluster, $\nu$ the increase of the correlation length, and $d_{\rm w}$ characterizes the mean-square displacement at the transition point. The emergence of subdiffusive  transport in the mean-square displacement and scaling behavior is accompanied by a vanishing diffusion coefficient as well as the appearance of an anomalous conductivity. Furthermore we have discussed the violation of the central limit theorem directly at the critical point and its consequences for the non-gaussian parameter.

The present study targets  ion transport in a quasi-frozen, three-dimensional environment. In these systems the disorder certainly contains correlations as the environment consists of some frozen-in configuration of a densely-packed strongly-interacting liquid. Thus it would be interesting to study transport also in disordered systems displaying some short-range correlations and see, whether the general scenario presented here qualitatively applies, in particular if the values of the \emph{dynamic} exponents remain unaffected by the correlations. Since it has been shown, that the value of the conductivity exponent is sensitive to the distribution of the weak conductances between different compartments of the clusters, the answer depends on how the local geometry affects the statistics of these bottlenecks and how they are tested by the particle's dynamics. Second, to make a comparison to realistic ion-conducting melts, one should take the finite lifetime of the local structures into account. Since  the obstacles slowly rearrange, once in a while blocked pockets will be connected to the infinite cluster and new dynamical processes enter the problem.

It would be interesting to study  also the dynamics on the infinite cluster for two-dimensional systems as minimal model for biological plasma membranes. Experiments on protein transport in the model cell membranes by fluorescence correlation spectroscopy (FCS)
have revealed anomalous transport~\cite{Gielen:2005,Weiss:2003,Avidin:2010}, which  has been interpreted as a manifestation of crowding.
Computer simulations modeling the obstructed motion in such membranes indeed observe anomalous transport~\cite{Saxton:1994,Saxton:2010,Sung:2006,Sung:2008,Sung:2008a,FCS_scaling:2010} and have identified the localization transition as a generic candidate to explain the experimentally observed findings.

\ack{
M.S and T.F.  acknowledge support via the DFG
research unit FOR-1394.
}

\small

\begin{thebibliography}{10}
\expandafter\ifx\csname url\endcsname\relax
  \def\url#1{{\tt #1}}\fi
\expandafter\ifx\csname urlprefix\endcsname\relax\def\urlprefix{URL }\fi
\providecommand{\eprint}[2][]{\url{#2}}

\bibitem{Goetze:Complex_Dynamics}
G\"{o}tze W 2009 {\em Complex Dynamics of Glass-Forming Liquids: A
  Mode-Coupling Theory\/} International Series of Monographs on Physics
  (Oxford: Oxford University Press)

\bibitem{Alba:2006}
Alba-Simionesco C, Coasne B, Dosseh G, Dudziak G, Gubbins K, Radhakrishnan R
  and Sliwinska-Bartkowiak M 2006 {\em J. Phys.: Cond. Matt.\/} {\bf 18}
  R15--R68

\bibitem{Mittal:2008}
Mittal J, Truskett T~M, Errington J~R and Hummer G 2008 {\em Phys. Rev.
  Lett.\/} {\bf 100} 145901

\bibitem{Scheidler:2000}
Scheidler P, Kob W and Binder K 2000 {\em Europhys. Lett.\/} {\bf 52} 277--283

\bibitem{Varnik:2002}
Varnik F, Baschnagel J and Binder K 2002 {\em Phys. Rev. E\/} {\bf 65} 021507

\bibitem{Nugent:2007}
Nugent C~R, Edmond K~V, Patel H~N and Weeks E~R 2007 {\em Phys. Rev. Lett.\/}
  {\bf 99} 025702

\bibitem{Lang:2010}
Lang S, Bo\ifmmode~\mbox{\c{t}}\else \c{t}\fi{}an V, Oettel M, Hajnal D,
  Franosch T and Schilling R 2010 {\em Phys. Rev. Lett.\/} {\bf 105} 125701

\bibitem{Lang:thesis}
Lang S 2010 {\em Glass transition in confined geometries described by
  mode-coupling theory\/} Master's thesis Johannes-Gutenberg Universit\"at
  Mainz

\bibitem{Krakoviack:2005}
Krakoviack V 2005 {\em Phys. Rev. Lett.\/} {\bf 94} 065703

\bibitem{Krakoviack:2007}
Krakoviack V 2007 {\em Phys. Rev. E\/} {\bf 75} 031503

\bibitem{Krakoviack:2009}
Krakoviack V 2009 {\em Phys. Rev. E\/} {\bf 79} 061501

\bibitem{Kurzidim:2009}
Kurzidim J, Coslovich D and Kahl G 2009 {\em Phys. Rev. Lett.\/} {\bf 103}
  138303

\bibitem{Kurzidim:2010}
Kurzidim J, Coslovich D and Kahl G 2010 {\em Phys. Rev. E\/} {\bf 82} 041505

\bibitem{Kim:2009}
Kim K, Miyazaki K and Saito S 2009 {\em Europhys. Lett.\/} {\bf 88} 36002

\bibitem{Goetze:1981a}
G{\"o}tze W, Leutheusser E and Yip S 1981 {\em Phys. Rev. A\/} {\bf 23} 2634

\bibitem{Goetze:1981b}
G{\"o}tze W, Leutheusser E and Yip S 1981 {\em Phys. Rev. A\/} {\bf 24} 1008

\bibitem{Goetze:1982}
G{\"o}tze W, Leutheusser E and Yip S 1982 {\em Phys. Rev. A\/} {\bf 25} 533

\bibitem{Ernst:1971a}
Ernst M~H and Weijland A 1971 {\em Phys. Lett. A\/} {\bf 34} 39

\bibitem{Stauffer:Percolation}
Stauffer D and Aharony A 1994 {\em Introduction to Percolation Theory\/} 2nd ed
  (London: Taylor \& Francis)

\bibitem{Kertesz:1983}
Kert{\'e}sz J and Metzger J 1983 {\em J.~Phys.~A\/} {\bf 16} L735--L739

\bibitem{Halperin:1985}
Halperin B~I, Feng S and Sen P~N 1985 {\em Phys. Rev. Lett.\/} {\bf 54}
  2391--2394

\bibitem{Machta:1985}
Machta J and Moore S~M 1985 {\em Phys. Rev. A\/} {\bf 32} 3164

\bibitem{Bruin:1972}
Bruin C 1972 {\em Phys. Rev. Lett.\/} {\bf 29} 1670--1674

\bibitem{Bruin:1974}
Bruin C 1974 {\em Physica\/} {\bf 72} 261

\bibitem{Alder:1983}
Alder B~J and Alley W~E 1983 {\em Physica A\/} {\bf 121} 523--530

\bibitem{Lowe:1993}
Lowe C~P and Masters A~J 1993 {\em Physica A\/} {\bf 195} 149--162

\bibitem{Lorentz_LTT:2007}
H{\"o}f\/ling F and Franosch T 2007 {\em Phys. Rev. Lett.\/} {\bf 98} 140601

\bibitem{Lorentz_VACF:2010}
Franosch T, H{\"o}f{}ling F, Bauer T and Frey E 2010 {\em Chem. Phys.\/} {\bf
  375} 540--547

\bibitem{Lorentz_2D:2010}
Bauer T, H{\"o}f{}ling F, Munk T, Frey E and Franosch T 2010 {\em Eur. Phys. J.
  Special Topics\/} {\bf 189} 103--118

\bibitem{Lorentz_PRL:2006}
H{\"o}f\/ling F, Franosch T and Frey E 2006 {\em Phys. Rev. Lett.\/} {\bf 96}
  165901

\bibitem{Lorentz_JCP:2008}
H{\"o}f\/ling F, Munk T, Frey E and Franosch T 2008 {\em J. Chem. Phys.\/} {\bf
  128} 164517

\bibitem{Lorentz_space:2010}
Franosch T, Spanner M, Bauer T, Schr\"oder-Turk G~E and H\"of{}ling F 2010 {\em
  J. Non-Cryst. Solids\/} {doi:} 10.1016/j.jnoncrysol.2010.06.051

\bibitem{FCS_scaling:2010}
H{\"o}f{}ling F, Bamberg K~U and Franosch T 2010 {\em Soft Matter\/} {doi:}
  10.1039/c0sm0718h, {arXiv:}1003.3762 [cond-mat.soft]

\bibitem{Percolation_EPL:2008}
Kammerer A, H{\"o}f\/ling F and Franosch T 2008 {\em Europhys. Lett.\/} {\bf
  84} 66002

\bibitem{Moreno:2006}
Moreno A~J and Colmenero J 2006 {\em J. Chem. Phys.\/} {\bf 125} 164507

\bibitem{Voigtmann:2009}
Voigtmann T and Horbach J 2009 {\em Phys. Rev. Lett.\/} {\bf 103} 205901

\bibitem{Lorentz_BSSM:2010}
Horbach J, Voigtmann T, H\"of{}ling F and Franosch T 2010 {\em Eur. Phys. J.
  Special Topics\/} {\bf 189} 141--145

\bibitem{Kikuchi:2007}
Kikuchi N and Horbach J 2007 {\em Europhys. Lett.\/} {\bf 77} 2600

\bibitem{Horbach:2002}
Horbach J, Kob W and Binder K 2002 {\em Phys. Rev. Lett.\/} {\bf 88} 125502

\bibitem{Meyer:2004}
Meyer A, Horbach J, Kob W, Kargl F and Schober H 2004 {\em Phys. Rev. Lett.\/}
  {\bf 93} 027801

\bibitem{Voigtmann:2006}
Voigtmann T and Horbach J 2006 {\em Europhys. Lett.\/} {\bf 74} 459--465

\bibitem{Lorentz:1905}
Lorentz H~A 1905 {\em Arch. N{\'e}erl. Sci. Exact Natur.\/} {\bf 10} 336--370

\bibitem{Elam:1984}
Elam W~T, Kerstein A~R and Rehr J~J 1984 {\em Phys. Rev. Lett.\/} {\bf 52} 1516

\bibitem{Spanner:thesis}
Spanner M 2010 {\em Transport in the correlated Lorentz model\/} Master's
  thesis Friedrich-Alexander-Universit\"at Erlangen N\"urnberg

\bibitem{benAvraham:DiffusionInFractals}
ben Avraham D and Havlin S 2000 {\em Diffusion and Reactions in Fractals and
  Disordered Systems\/} (Cambridge: Cambridge University Press)

\bibitem{Felderhof:2005}
Felderhof B 2005 {\em Journal of Physical Chemistry B\/} {\bf 109} 21406--21412

\bibitem{Jeney:2008}
Jeney S, Lukic B, Kraus J~A, Franosch T and Forro L 2008 {\em Phys. Rev.
  Lett.\/} {\bf 100} 240604

\bibitem{Franosch:2009}
Franosch T and Jeney S 2009 {\em Phys. Rev. E\/} {\bf 79} 031402

\bibitem{Kubo:Statistical_Physics_II}
Kubo R, Toda M and Hashitsume N 1991 {\em Statistical Physics II,
  Nonequilibrium Statistical Mechanics\/} (Berlin, Heidelberg: Springer)

\bibitem{Tuck:1967}
Tuck E 1967 {\em Math. Comput.\/} {\bf 21} 239--241

\bibitem{Gielen:2005}
Gielen E, Vercammen J, S\'{y}kora J, Humpol{\'i}{\v{c}}kov{\'a} J, van de Ven M,
  Benda A, Hellings N, Hof M, Engelborghs Y, Steels P and Ameloot M 2005 {\em
  C. R. Biologies\/} {\bf 328} 1057--1064

\bibitem{Weiss:2003}
Weiss M, Hashimoto H and Nilsson T 2003 {\em Biophys. J.\/} {\bf 84} 4043--4052

\bibitem{Avidin:2010}
Horton M~R, H\"of{}ling F, R\"adler J~O and Franosch T 2010 {\em Soft Matter\/}
  {\bf 6} 2648--2656

\bibitem{Saxton:1994}
Saxton M~J 1994 {\em Biophys. J.\/} {\bf 66} 394--401

\bibitem{Saxton:2010}
Saxton M~J 2010 {\em Biophys J\/} {\bf 99} 1490--1499

\bibitem{Sung:2006}
Sung B~J and Yethiraj A 2006 {\em Phys. Rev. Lett.\/} {\bf 96} 228103

\bibitem{Sung:2008}
Sung B~J and Yethiraj A 2008 {\em J.~Chem. Phys.\/} {\bf 128} 054702

\bibitem{Sung:2008a}
Sung B~J and Yethiraj A 2008 {\em J. Phys. Chem. B\/} {\bf 112} 143--149

\end{thebibliography}

\providecommand{\newblock}{}

\end{document}